\documentstyle[11pt]{article}
\input epsf

\title{\bf Rapidity and centrality dependence in the
percolating colour strings scenario}
\author{M.A.Braun$^{a,b}$ and C.Pajares$^a$\\
$^a$ Dep. of Elementary Particles,\\
Univ. of Santiago de Compostela, 15706,
Santiago de Compostela, Spain, \\
$^b$ Dep. of High Energy physics,
 University of S.Petersburg,\\
198504 S.Petersburg, Russia}
\def\beq{\begin{equation}}
\def\eeq{\end{equation}}
\def\noi{\noindent}
\begin{document}
\maketitle
\medskip

\noi {\bf Abstract}

In AA collisions fusion and percolation of colour strings is
studied at fixed rapidity $y$. Distribution of strings in rapidity
is obtained from the observed rapidity spectra in pp collisions.
For  $y$-dependence of multiplicities in Au-Au collisions
good agreement is obtained with the existing experimental data.
Predictions for LHC energies coincide with the extrapolation of the
data. Agreement with the data of the transverse momentum spectra
requires introduction of quenching into the model.

\section{Introduction}
The color string model with fusion \cite{Ref1}, \cite{Ref2} and percolation
\cite{Ref3}-\cite{Ref6} has produced results on multiplicities of secondaries
which are in general agreement with the existing experimental data. The
string
fusion model predicted a strong reduction of multiplicities both at
RHIC and
LHC energies. At the time of the ALICE Technical proposal of 95 \cite{Ref7},
most of model predictions, (including VENUS \cite{Ref8}, HIJING
\cite{Ref9},
SHAKER \cite {Ref10} and DPM \cite{Ref11}) for LHC energies were more
than 4000
charged particles at central rapidity region for central $(b\leq 3\ fm)$
Pb-Pb
collisions and only the prediction of the string fusion model was much lower.
Since then many of the models have lowered their predictions introducing
several
mechanisms, such as the triple pomeron coupling in DPM \cite{Ref12},
stronger
shadowing in HIJING \cite{Ref13} or other modifications in VENUS \cite{Ref14}.
In the parton saturation picture, predictions for
the central
rapidity density of charged particles per participant for central Pb-Pb at
 the LHC
energy range from around 15 \cite{Ref15} to a lower value around 9
 \cite{Ref16}. Assuming that the observed
geometrical
scaling for the saturation momentum in lepton-hadron scattering
is also valid for the nucleus-nucleus scattering, the value
around 9 is also obtained \cite{Ref17}.
In percolation of strings the obtained
values  7.3 \cite{Ref18} and 8.6 \cite{Ref19} are not far from the above
ones,
as expected, given the similarities between percolation of strings and
saturation of partons \cite{Ref20}. In any case, these values
lie above 6.4 which is
obtained by extrapolation to the LHC energy of the values
experimentally found at
$\sqrt{s}=19.4, 62.8, 130$ and 200 GeV \cite{Ref21}. The values obtained in
percolation have some uncertainty due to  simplifications done in the
calculations, mainly related to the dependence of the
multiplicity on the energy of a simple string and on the rapidities of
fused
strings. One of the goals of this paper is to take into account
both these dependencies, using
as an input the rapidity and energy dependence of
multiplicities in pp
collisions. We try to answer whether the charge particle density per
participant
is compatible with the one obtained in the mentioned extrapolation.
We assume
that strings occupy different regions in the available rapidity interval.
Then at
different rapidities one will see different number of overlapping strings,
depending on the rapidities of the string ends. As a result the string
density
and the process of percolation become dependent on rapidity together
with all
the following predictions for observable quantities. The central  part of our
derivation is the calculation of the number of strings at a fixed rapidity
which
follows from the known parton distributions in the projectile and target. To
this aim we shall use the simple parton distributions employed in the review
\cite{Ref11}. Also, we pay attention to the fragmentation rapidity region. As
an input we take the observed limiting fragmentation scaling in pp
collisions.
We then obtain  a similar scaling for A-A collisions, in  agreement with
the experimental data \cite{Ref22}.

One of the most interesting features of the
RHIC
data is the suppression of high transverse momenta. The nuclear
modification
factor defined as the ratio between inclusive A-A cross section normalized to
the number of collisions and the inclusive proton-proton cross-section
is found to lie below unity,
in disagreement with the perturbative QCD expectations.

In the previous paper
\cite{Ref6}, in the framework of percolation of strings,
a reasonable agreement with the data was obtained,
describing A-A collisions as an exchange of
clusters of overlapping strings.
In percolation each cluster behaves like a new
string with a larger tension, its value depending on the number of
strings fused into the cluster and the cluster's transverse area.
Fragmentation of each cluster was assumed to
give rise
to an exponential distribution in $p^2_T$. Superposition of
different exponential distributions then builds up a power-like
distribution $\sim p_T^{-\kappa}$
with $\kappa$ inversely related to  the magnitude of the dispersion in the
number of different clusters. At low density, there is no overlapping of
strings, thus no fluctuations and $\kappa$ is large.
As the density increases,
so does the string overlapping and more clusters are formed
with different number  of
strings. So the dispersion increases and $\kappa$ decreases.
Finally, at very high
density, above the percolation threshold, there remains a single large
cluster of nearly all the
strings. Therefore, there are no fluctuations and $\kappa$ increases again
becoming large. In this way, the suppression of high $p_T$ at large density
follows as a result of  formation of large clusters of color strings.
In ~\cite{Ref6} we  assumed that the spectrum of a simple string
was a single exponential in $p^2_T$. However, at low string  density,
as in  pp collisions, when fusion of strings is insignificant,
the  experimental data clearly show a power-like
tail for the $p_T$ distribution.

In this paper we study  this point with more attention.
Instead of the exponential distribution we take the  standard
power-like parameterization for pp collisions \cite{Ref23}.
The resulting $p_T$
distribution for central A-A collisions is again found
suppressed at high $p_T$ but not
enough to agree with the data. In order to describe the data we need
additional
suppression which would physically correspond to the fact that the
produced particle,
passing through a large cluster and interacting with the strong
chromoelectric
field, looses a part of its energy. This result is not unexpected.
In fact a
version of HIJING \cite{Ref24}, with a string junction and
doubling the string
tension to simulate stronger color-fields, is able to explain the
difference between baryons and mesons in the low and mid $p_T$ range
but at high $p_T$ some jet quenching mechanism is needed.
In our framework quenching at high $p_T$ may be introduced in a simple
phenomenological manner by taking the average $p_T^2$ of a cluster of $n$
strings to grow with $n$ more slowly than $\sqrt{n}$, as predicted
for a single small cluster in absence of others. Choosing an appropriate
$n$ dependence of the average $p_T^2$ for clusters at a given string
density allows to obtain a reasonable agreement with the experimental
data on $p_T$ dependence in A-A collisions.

\section{pp collisions}
\subsection{Multiplicities and numbers of strings}
The starting point for the calculation of fusion and
percolation of strings in heavy -ion collisions is the
distribution of strings in proton-proton collisions, where
effects of fusion and percolation are very small.
Our strategy will be to extensively use the existing
experimental data for the multiplicity per unit
rapidity $d\mu^{pp}(y)/dy\equiv \mu^{pp}(y)$ in
pp collisions to extract
the necessary distribution of strings in $y$ from them.

We recall that in the original DPM model without
string fusion the multiplicity is given by a sum of
contribution from the strings formed in the collision.
In particular, in a configuration with
$n=2k$ strings formed,
which corresponds to the exchange of $k$ pomerons ~\cite{Ref11},
the multiplicity is given by
\beq
\mu^{pp}_{n}(y)=\int_{-Y/2}^{Y/2}
\prod_{i=1}^ndu_idw_i p(u_1,...,u_n)
t(w_1,...,w_n)\sum_{j=1}^n\mu_j(y,u_j,w_j).
\eeq
We work in the c.m system of colliding protons;
 $Y$ is the overall rapidity admissible for nearly massless
quarks.
It is related to the beam rapidity as
\beq
Y=Y_{beam}+\ln\frac{m^2}{\mu^2},
\eeq
where $m$ is the nucleon mass and $\mu$ the quark average transverse mass.
The strings are enumerated according to their flavour
content, that is according to which quark they are attached.
Number 1 corresponds to the quark-diquark (qd)
string, number 2 to  diquark-quark (dq) string and all the rest
correspond to sea quarks, which include
$s\bar{s}$ and $\bar{s}s$ strings. Ends of  strings in rapidity
are denoted by $u_i$ in the projectile and $w_i$ in the target.
Distributions $p(u_1,...,u_n)$ and $t(w_i,...w_n)$ give the
probability to find the relevant quarks in the projectile and
target proton respectively. Note that in the assumed notation
$p(u_1,...u_n)$ gives the probability to find in the proton
the valence quark at rapidity $u_1$ the diquark at rapidity $u_2$
and the sea quarks at rapidities $u_3,...u_n$. Distribution
$t(w_1,...w_n)$, on the other hand, gives the probability to find
the diquark at rapidity $w_1$, the quark at rapidity $w_2$ and sea
quarks at rapidities $w_3,...w_n$.
Function  $\mu_j(y,u,w)$ gives the multiplicity per unit
rapidity from the $j$th string at rapidity $y$ provided its
ends are at $u$ in the projectile and $w$ in the target.
Obviously this probability is zero if the string lies outside
rapidity $y$. So $\mu_j(y,u,w)$ has a form
\beq
\mu_j(y,u,w)=\rho(y,u,w)\tilde{\mu}_j(y,u,w),
\eeq
where
\beq
\rho(y,u,w)=\theta(u-w-y_0)\theta(u-y)\theta(y-w)
+\theta(w-u-y_0)\theta(w-y)\theta(y-u).
\eeq
Here the two terms correspond to the two possibilities
of the higher rapidity  end of the string to lie  on the projectile or
on the target parton. The rapidity interval $y_0$ corresponds to the
minimal extension of the string in rapidity. We take $y_0=2$.

The string density $dN^{pp}_{n}(y)/dy\equiv
N^{pp}_{n}(y)$ at a given
rapidity is given by an expression similar to (1) but without
$\tilde{\mu}$:
\beq
N^{pp}_{n}(y)=\int_{-Y/2}^{Y/2}\prod_{i=1}^ndu_idw_i p(u_1,...,u_n)
t(w_1,...,w_n)\sum_{j=1}^n\rho(y,u_j,w_j).
\eeq
To analyse fusion probabilities in nuclear collisions we need
to know the latter quantity.

In principle, knowledge of  the distributions $p(u_1,...,u_n)$
and $t(w_1,...,w_n)$ and of string luminosities $\tilde{\mu}_j(y,u,w)$
allows to calculate both $\mu_n^{pp}(y)$ and $N_n^{pp}(y)$. This was done in the
extensive calculations within the original DPM model ~\cite{Ref11}.
However these input quantities are in fact poorly known and our idea
is to directly relate $\mu^{pp}(y)$ and $N^{pp}(y)$ using the experimental
data for the former. To do this we assume that string luminosities
 are approximately $y$-independent and the same
for all type of strings:
\beq
\tilde{\mu}_j(y,u,w)=\mu_0.
\eeq
This approximation has been widely used in
analytical studies of string fusion. It can be justified for relatively
long strings far from their ends, when particle production can be well
described by the Schwinger mechanism of pair creation in a strong field.
With strings of finite dimension it may be considered as a sort of
averaging over their length and rapidity of emission.
With this approximation we obtain a simple and direct relation
between the multiplicity and number of strings per unit rapidity:
\beq
\mu^{pp}_{n}(y)=\mu_0N^{pp}_{n}(y).
\eeq
In fact this relation for the central region was extensively used
in earlier studies of string percolation.

Relation (7) allows to find only the total number of strings per
unit rapidity from the experimental data on multiplicities. However
we need something more. In nucleus-nucleus collisions separately
enter multiplicities coming from the valence strings and sea strings.
Obviously one cannot find each of them from the experimental data.
So we choose to calculate the contribution of valence strings from
the theoretical formulas (5) and (7) and then, subtracting this
contribution from the experimental multiplicities, find the
contribution from sea strings. This procedure can be justified
by the fact that distributions of valence quarks are much better
known and less dependent on the overall energy than the sea
contribution.

\subsection{Total and sea strings from the experimental data}
We calculate the number of quark-diquark strings from
(5) as
\[
N^{qd}_{n}(y)=
\int_{-Y/2}^{Y/2}dudwq_n^{(p)}(u)d_n^{(t)}(w)\rho(y,u,w)
\]\beq=
\int_{y}^{Y/2}du\int_{-Y/2}^{w_1}dw
\Big(q_n ^{(p)}(u)d_n^{(t)}(w)+d_n^{(t)}(u)q_n^{(p)}(w)\Big),\ \
 w_1=\min\{y,u-y_0\}.
\eeq
Here $q_n ^{(p)}(u)$ and $d_n^{(t)}(w)$ are inclusive probabilities to find
a valence quark in the projectile and a diquark in the target at
rapidities $u$ and $w$ respectively in a configuration with $n$
strings. The second
term in (8) corresponds to
inverse strings whose upper ends lie on the target diquark.
In our symmetric case the number of diquark-quark strings is
obviously the same, so that the total number of valence strings is
just twice the expression (8).

The final number of valence strings at given $y$ is obtained after
averaging over
the number of formed strings:
\beq
N^{v}(y)=2\sum_{k=1}\omega_kN^{qd}_{(2k)}(y)\equiv
2\langle N^{qd}_{(n)}(y)\rangle.
\eeq
Here $\omega_k$ is the probability for the exchange of $k$
pomerons, given by ~\cite{Ref11}
\beq
\omega_k=\frac{\sigma_k}{\sum_{l=1}\sigma_l},
\eeq
where $\sigma_k$ is the cross-section for $k$ inelastic collisions.
It is standardly taken in the K.A.Ter-Martirosyan model ~\cite{Ref25}
\beq
\sigma_n(s)=2\pi\int_0^{\infty}bdbe^{-2\chi}\frac{(2\chi)^n}{n!},
\eeq
where the eikonal $\chi$ corresponds to the single pomeron exchange
\beq
\chi(s,b)=C(s)e^{-b^2/b_0^2(s)},
\eeq
with
\beq
b_0^2(s)=4R_N^2+4\alpha'\left(\ln s-i\frac{\pi}{2}\right),\ \
C(s)=\frac{g^2}{b_0^2(s)}\Big(se^{-i\pi/2}\Big)^{\alpha-1},
\eeq
$\alpha$ and $\alpha'$ are the pomeron intercept and slope,
$g$ is its coupling to the proton and $R_N$ the proton radius.
Some improvement of these $\sigma_k$ to include the triple
pomeron interaction and diffractive states may be found in ~\cite{Ref11}.

To calculate the number of valence strings per unit rapidity
we have to know
the inclusive distributions of quarks and diquarks.
Following ~\cite{Ref11} we choose the exclusive distribution
$p(u_1,u_2,...u_n)$  for a projectile in a factorized
form
\beq
p_n(u_1,u_2,...u_n)=c_n\delta(1-\sum_{i=1}^nx_i)\prod_{i=1}^n x_i^{\mu_i},
\eeq
where for the quark $\mu_1=1/2$ and for the diquark $\mu_2=5/2$
For the sea quarks and antiquarks we take
$\mu=1/|\ln x_c|$,
where in accordance with ~\cite{Ref11} $x_c=m_c/\sqrt{s}$
with $m_c=0.1$ GeV
 is a  cutoff at small $x$.
Scaling variables  are related to
rapidities as
\beq
x=e^{-Y/2+u},
\eeq
Note that the distributions $p_n(u_1,u_2,...u_n)$ are defined and
normalized in the interval $0<x<1$, that is for $-\infty<u<Y/2$.
The actual strings are formed only in the part of this interval
with $u>-Y/2$. This circumstance is inessential for valence quarks
whose distributions  rapidly vanish towards small values of $x$.
For the distributions
in the target one has only to invert the rapidities $u\to -w$ in (15).
Integration over the scaling variables of unobserved partons
gives the desired inclusive
distributions. For the valence quark we find
\beq
q_p^{(n)}(x)=
c_vx^{1/2}(1-x)^{3/2+(n-2)\mu},\ \
c_v=\frac{\Gamma(3+(n-2)\mu}{\sqrt{\pi}\Gamma(5/2+(n-2)\mu)}.
\eeq
For the diquark
\beq
d_p^{(n)}(x)=c_d
x^{5/2}(1-x)^{-1/2+(n-2)\mu},\ \ c_d=
\frac{4\Gamma(3+(n-2)\mu}{3\sqrt{\pi}\Gamma(1/2+(n-2)\mu)}.
\eeq

After the averaged valence string number is found according to (9)
we have to transform it into the valence multiplicity
using (7). The value of $\mu_0$ can be found from the observed
plateau hight assuming that at $y=0$ all strings contribute.
Their average number  can be found
from (9) as $N^{pp}=2\langle k\rangle$.
As a result we find values of $\mu_0$ slowly rising with
energy and visibly saturating at TeV energies. In Fig. 1
we show these values extracted from the data  ~\cite{Ref23}
together with their extrapolation to the LHC energies in the
assumption that the plateau in the pp multiplicity
distribution rises linearly with $\ln s$.
%
\begin{figure}[ht]
\centerline{\epsfbox{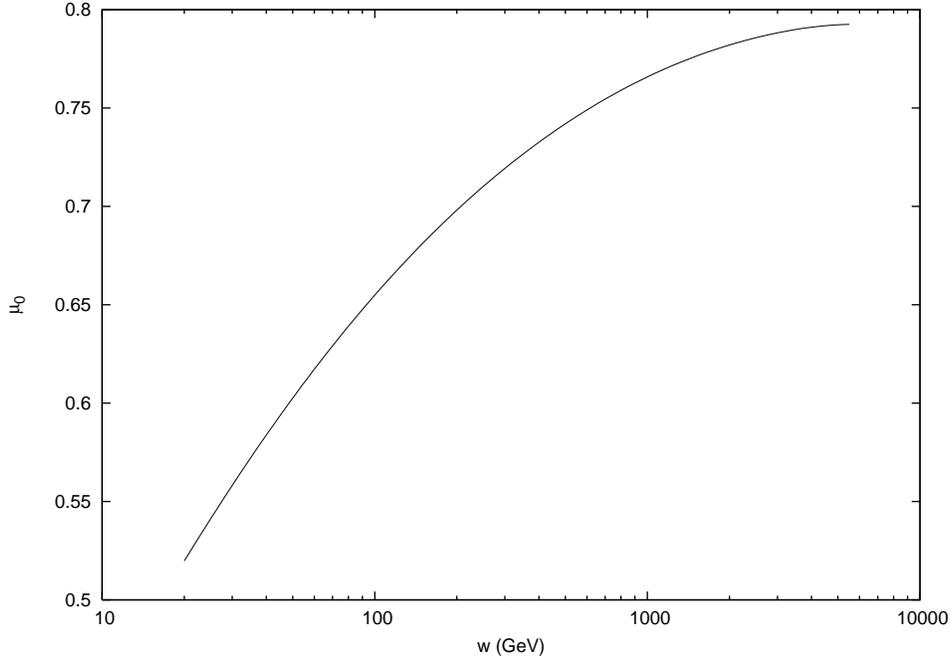}}
\caption{String luminosity as a function of energy}
\label{Fig1}
\end{figure}
%
The obtained multiplicities from valence strings
vanish in the fragmentation region too slowly as compared
to the experimental data, and at $y-Y_{beam}>0$ become
greater than the latter. This is obviously related to our
assumption of  a constant string luminosity throughout the
string length, whereas it should go to zero at its ends.
Put in other words, in our approach the total energy is
conserved in its division between different strings due to the
$\delta$-function in (14) but it is not conserved inside each
separate string, since near the string end its luminosity should vanish.
To cure this defect in a simple manner we just assume that as soon as the
calculated valence contribution becomes larger that the data
we substitute the former by the latter, assuming that in this deep
fragmentation region sea strings do not contribute at all.

With thus obtained valence contribution to the multiplicities we
find the sea contribution just as the difference between the total
and valence one. Dividing it by $\mu_0$ we find the number of sea
strings per unit rapidity in pp collisions $N^{s,pp}(y)$. In fact we need
not exactly this number
but the one in the assumption that all strings are of the sea type,
which is obtained from it by rescaling
\beq
N^{s}(y)= \frac{N^{pp}}{ N^{pp}-2}
N^{s,pp}(y).
\eeq
This quantity is shown in Fig. 2 for different values of the
collision energy $\sqrt{s}$.
\begin{figure}[ht]
\centerline{\epsfbox{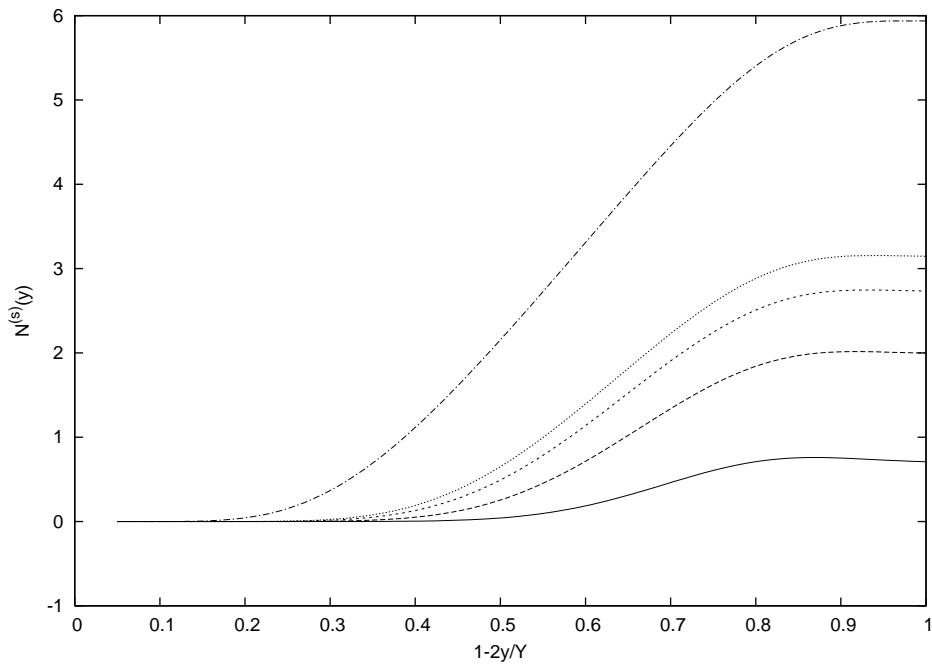}}
\caption{Sea string contribution to multiplicities per
nucleon as a function of rapidity.
Curves from bottom to top correspond to energies
19.4, 62.8, 130, 200 and 6000 GeV.}
\label{Fig2}
\end{figure}

\section{hA and AA collisions}
Generalization to hA and AA collisions is straightforward and
follows ~\cite{Ref11}. At fixed impact parameter $b$
one introduces the average numbers of participants $2\nu_A(b)$ and
collisions
$\nu(b)$. Then the number of strings in AA collisions at given $b$ and $y$ is
\beq
N_{AA}(b,y)=\nu_{par}(b) N^{pp}(y)+\Big(\nu_{col}(b)-\nu_{par}(b)\Big)
N^s(y).
\eeq
Here $N^{pp}(y)$ is obtained from the observed multiplicity in pp collisions
according to (7) and $N^s(y)$ is given by (18).

One can easily further generalize (19) to collisions of different nuclei
(see ~\cite{Ref11}).

In the Glauber approach the numbers $\nu_{par}$ and $\nu_{col}$ for
AA collisions are obtained as follows
\beq
\nu_{par}(b)=A\frac{\int d^2b'T_A(b')\Big(1-e^{-A\sigma T_A(b-b')}\Big)}
{1-e^{-A^2\sigma T_{AA}(b)}},\ \
\nu_{col}(b)=A^2\frac{\sigma T_{AA}(b)}{1-e^{-A^2\sigma T_{AA}(b)}},
\eeq
where $\sigma$ is the total pp-cross-section, $T_A(b)$ is the nuclear
profile function normalized to unity and
\beq
T_{AA}(b)=\int d^2b'T_A(b')T_A(b-b').
\eeq
For hA collision, as mentioned, $\nu_{par}=1$ and
\beq
\nu_{col}(b)=A\frac{\sigma T_A(b)}{1-e^{-A\sigma T_A(b)}}.
\eeq
Note that for AA collisions in the above formulas the denominator is written
in the so-called optical approximation ~\cite{opt}. As is well-known,
it works reasonably
well except close to the nucleus boundary, where the collision numbers
obtained
from (20) may be quite deceptive.

It is customary to take the profile
function $T_A(b)$ generated by the Woods-Saxon nuclear density. However for
our purpose it is more convenient to assume the nucleus to have a well-defined
radius, which allows to determine the interaction area in the transverse plane
as just the area of the overlap. For this reason we take the nucleus as a
sphere of radius $R_A= A^{1/3}\cdot 1.2$ fm, which gives
\beq
T_A(b)=\theta(R_A-b)\frac{2\sqrt{R_A^2-b^2}}{V_A},
\eeq
where $V_A$ is the nuclear volume. With this choice the most peripheral
collisions occur at $b=2R_A$ with $\nu_{col}=\nu_{par}=1$.

\section{Multiplicities and $p_T$ distributions}
In our previous studies of string fusion
we always stressed that it  can only
occur in the common rapidity interval.
However our attention was mostly centered on the central rapidity region
where all (or nearly all) strings contribute, so that the requirement
of common rapidity interval was of no relevance and strings could be
considered  as of practically infinite length in rapidity.
Now we study the fusion process in more detail. At a fixed rapidity
$y$ only strings which pass through this rapidity can fuse.
Formulas of the previous sections allow to find the original number
of strings $N(b,y)$ stretched  between the projectile and target at fixed
rapidity layer $y$ and impact parameter $b$. According to the percolation
colour strings scenario these strings in fact fuse into strings with higher
colour. The intensity of fusion is determined by the dimensionless
percolation parameter $\eta$ proportional to the string density in the
interaction area
\beq
\eta(b,y)=\frac{N(b,y)s_0}{S(b)},
\eeq
where $s_0=\pi r_0^2$ is the transverse area of the string and $S(b)$
is the interaction area, that is, the overlap area in case of AB collisions.
Obviously in our case the percolation parameter depends both on $b$
and $y$. At $\eta\sim 1.2\div 1.3$ fusion of strings leads to their
percolation  and formation of macroscopic string clusters. This
phenomenon will take part only in restricted intervals of $b$ and $y$,
predominantly at central collisions and rapidities, where
the effects of string fusion and percolation will be most noticeable

Considering the case of AA collisions at reasonably high energies we
shall assume the total number of strings high enough to allow use
of the thermodynamic limit, in which the total areas of $n$-fold fused
strings $S_n$ become distributed according to the Poisson law with
$\langle n\rangle =\eta(b,y)$:
\beq
S_n(b,y)=S(b)e^{-\eta(b,\nu)}\frac{\eta^n(b,\nu)}{n!}.
\eeq
Due to averaging of the direction of colour, the $n$-fold fused string
emits the number of particles which is only $\sqrt{n}$ times greater
that the simple string. So the total production rate at fixed $b$ and $y$
will be given by
\beq
\mu(b,y)=\mu_0\frac{S(b)}{s_0}e^{-\eta(b,\nu)}\sum_{n}\sqrt{n}
\frac{\eta^n(b,\nu)}{n!},
\eeq
where $\mu_0$ is the production rate from the  single string, which,
as stated above, we
assume to be independent of $y$ but dependent on energy.

As to the $p_T$ distribution, we use a slightly generalized
model introduced and discussed in
~\cite{Ref26}, in which the normalized probability $w_n(p)$
to find a particle with transverse momentum $p$ emitted from the
$n$-fold fused string is given by
\beq
w_n(p)=\frac{(\kappa_n-1)(\kappa_n-2)}{2\pi p_n^2}
\Big(\frac{p_n}{p+p_n}\Big)^{\kappa}.
\eeq
Here for $n=1$ the parameters are determined by the experimental
data on pp collisions:
\beq
p_1=2\ {\rm GeV/c},\ \ \kappa_1=19.7-0.86\ln E_{cm}
\eeq
and $E_{cm}$ is the c.m. energy in GeV.
With $n>1$ from the string fusion scenario it follows that the average
transverse momentum squared  of the particles emitted from the $n$-fold
fused string is $n^{1/2}$ greater than for a single string. This gives
a relation between $p_n$ and $\kappa_n$
\beq
p_n^2=n^{1/2}p_1^2\frac{(k_n-3)(k_n-4)}{(k_1-3)(k_1-4)}.
\eeq
So the distribution from $n$-fold string is fully determined by the
$n$-dependence of $\kappa_n$. In ~\cite{Ref26} the simplest choice of
$n$-independent $\kappa_n$ was used.
However this simple choice does not allow to obtain the $p_T$
dependence in agreement with the data at RHIC.
To improve our description, we introduce corrections to the original
string picture which correspond to non-linear phenomena in string
clustering and influence both $\kappa_n$ and the behaviour of $p_n$

In fact the value of $\kappa$ controls the difference of the distribution
from the purely exponential one, passing into the latter at very large
$\kappa$. For a single string a finite $\kappa_1$ may be thought of
as a result of fluctuations in the string tension (or equivalently
its transverse area) ~\cite{Ref27}. One may expect this fluctuations to
grow as many
string
fuse and so the value of $\kappa_n$ should fall with $n$. However there is
another effect acting in the opposite direction. As many strings fuse into
clusters, multiple interactions of emitted particle inside the clusters
should lead to thermalization of the particle spectra making it closer to
an
exponential. Thus eventually at large $n$ parameter $\kappa_n$ should grow
to large values. Naturally we cannot determine the exact form of
the $n$ dependence of $\kappa_n$ from purely theoretical reasoning.
We can only think that the change from fall to growth should occur in the
vicinity of the percolation threshold and that in any case $\kappa$ cannot be
smaller than 4 to have a convergent $<p^2>$. In practice we
take $\kappa_n$ as
\beq
\kappa_n=\kappa_1+a(n-1)+b(n-1)^2
\eeq
and try to adjust
$a$ and $b$ to get a better agreement with the experimental
data. In fact the results are not very sensitive to the choice of $a$ and $b$
provided they are taken to have a behaviour of
$\kappa_n$ in agreement with the above general theoretical observations.
However the generalization (30) is not sufficient to bring our predictions
in agreement with the observed quenching of the ratios $R_{AA}$ at RHIC.
To this aim we have to introduce some quenching also in the string
picture at large values of $\eta$. It corresponds to the fact that passing
through a large cluster volume and interacting with the strong
chromoelectric field  the
produced particles  loose a part of their energy ~\cite{Ref28}. On our
phenomenological level it would correspond to the behaviour of the average
transverse momentum squared as
\beq
<p^2>_n=n^{\alpha_n}<p^2>_1
\eeq
with the exponent $\alpha_n$ less than 1/2 and diminishing with $n$.
Similarly to (30) we parameterize
\beq
\alpha_n=\frac{1}{2}+c(n-1)+d(n-1)^2.
\eeq
The comparison with the experimental data determines the
optimal fit for the  parameters $a$, $b$, $c$ and $d$. The resulting
values for $\kappa_n$ and  $\alpha_n$ are shown in Figs. 3 and 4.
\begin{figure}[ht]
\centerline{\epsfbox{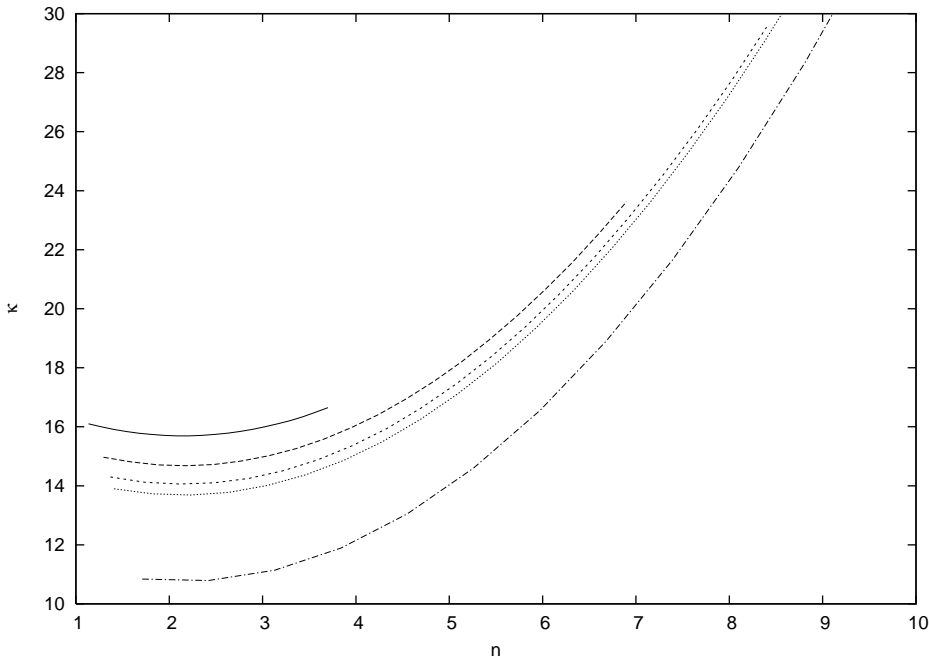}}
\caption{Parameter $\kappa_n$ in the distribution (27) for the $n$-fold
fused string. Curves from top to bottom correspond to energies
19.4, 62.8, 130, 200 and 6000 GeV. The curves end at  maximal $n$ reached
at these energies}
\label{Fig3}
\end{figure}
\begin{figure}[ht]
\centerline{\epsfbox{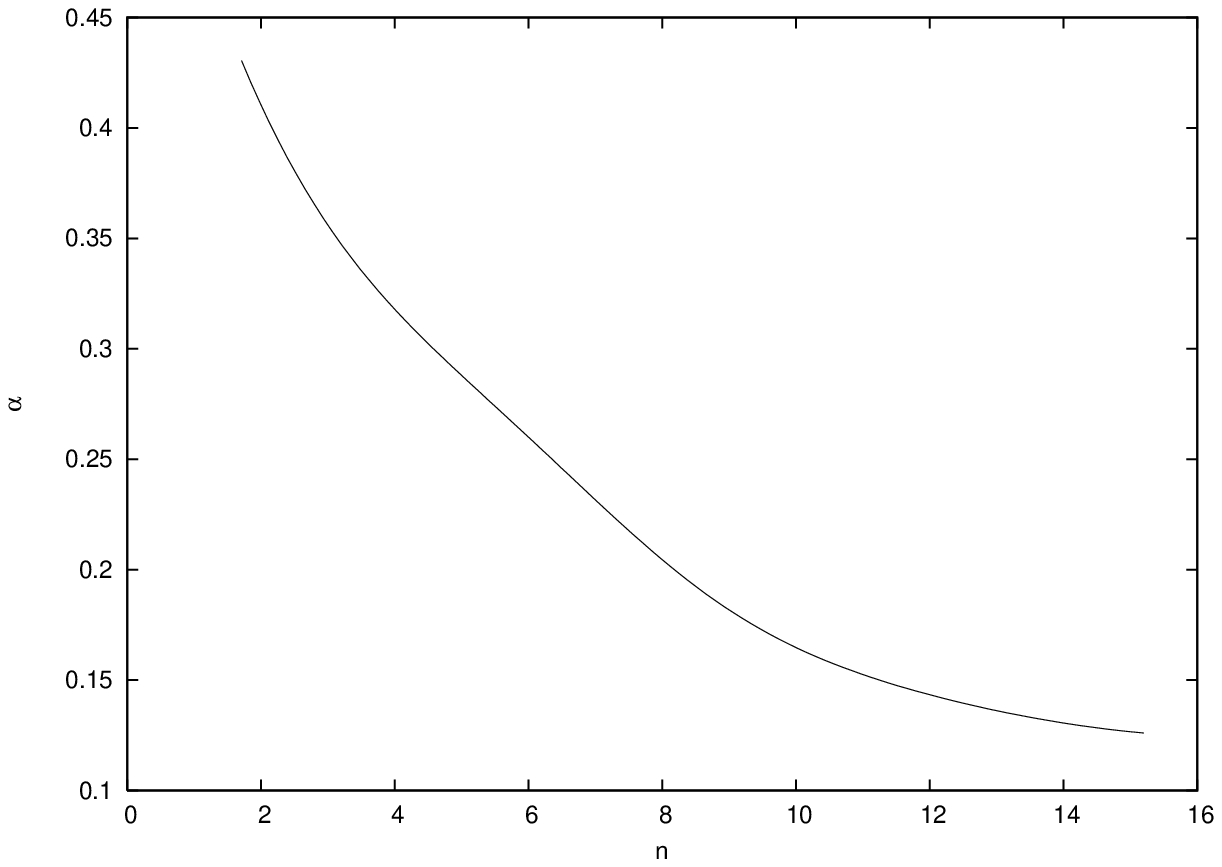}}
\caption{Power $\alpha_n$ for the $n$ dependence of the average transverse
momentum (31) of the $n$-fold fused string}
\label{Fig4}
\end{figure}

Averaging with the distribution in $n$ we find the final distribution
in $p$ from the fusing strings at  given $b$ and $y$ as
\beq
w(p,b,y)=a(b,y)\sum_{n=1}w_n(p)\frac{\eta^n(b,\nu)}{n!},\ \
a^{-1}(b,y)=e^{\eta(b,y)}- 1
\eeq
(the change in the normalization is due to the restriction $n\geq 1$).

\section{Numerical results}

We studied Au-Au collisions at energies 19,4, 62.8, 130, 200 and 6000 GeV
corresponding to the existing experimental data and expected at LHC.
Using our results on the string numbers in pp collisions we
calculated their numbers  in nucleus-nucleus collisions at a given $y$.
Knowing these numbers and also numbers of participants and collisions
we then determined  values of the percolation parameter
$\eta(y,b)$ at different rapidities and impact parameters.
We have taken the transverse radius of the single string 0.3 fm.
In Fig. 5 and 6  we illustrate  values of $\eta(y,b)$ as a function of $y$
for central collisions
and as a function of $b$ at mid-rapidity.
\begin{figure}[ht]
\centerline{\epsfbox{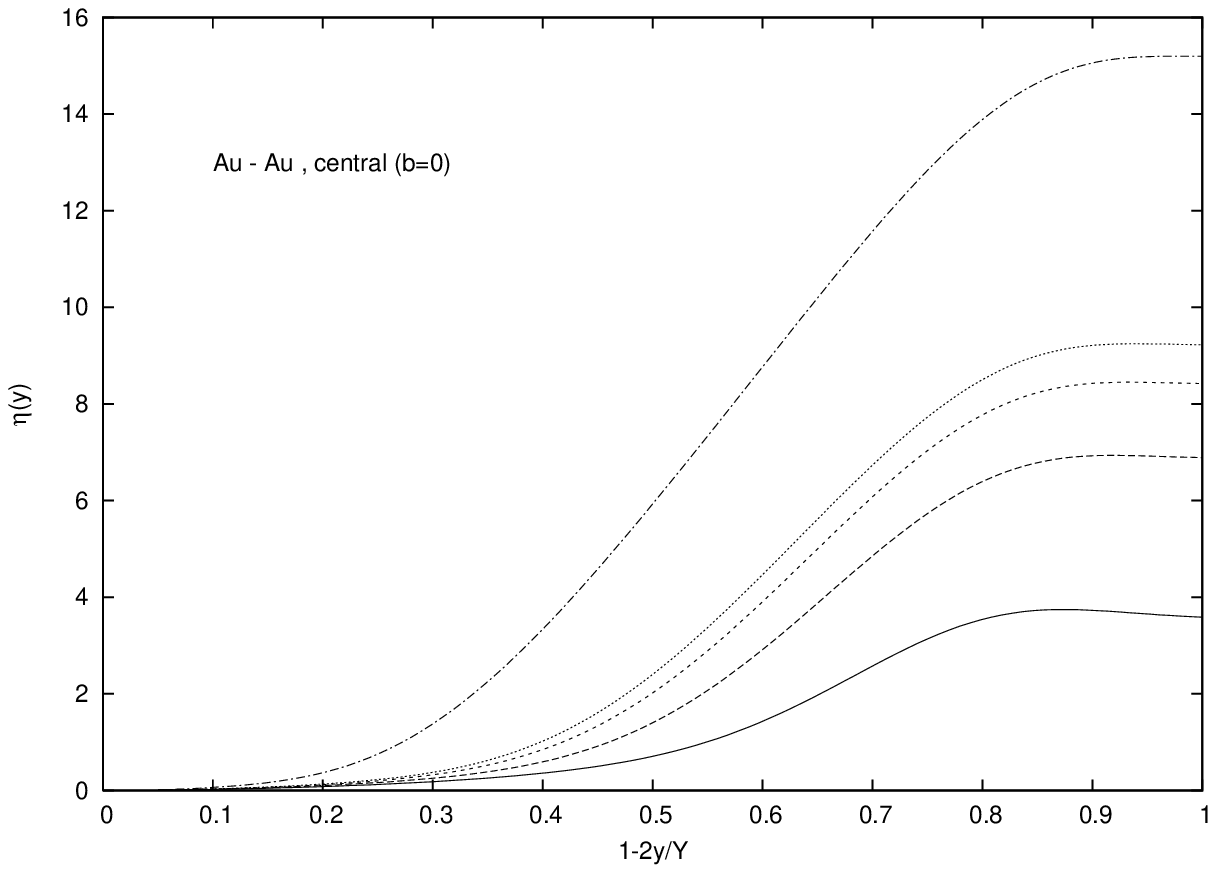}}
\caption{Percolation parameter $\eta$ in central
Au-Au collisions as a function of rapidity.
Curves from bottom to top correspond to energies
19.4, 62.8, 130, 200 and 6000 GeV.}
\label{Fig5}
\end{figure}
\begin{figure}[ht]
\centerline{\epsfbox{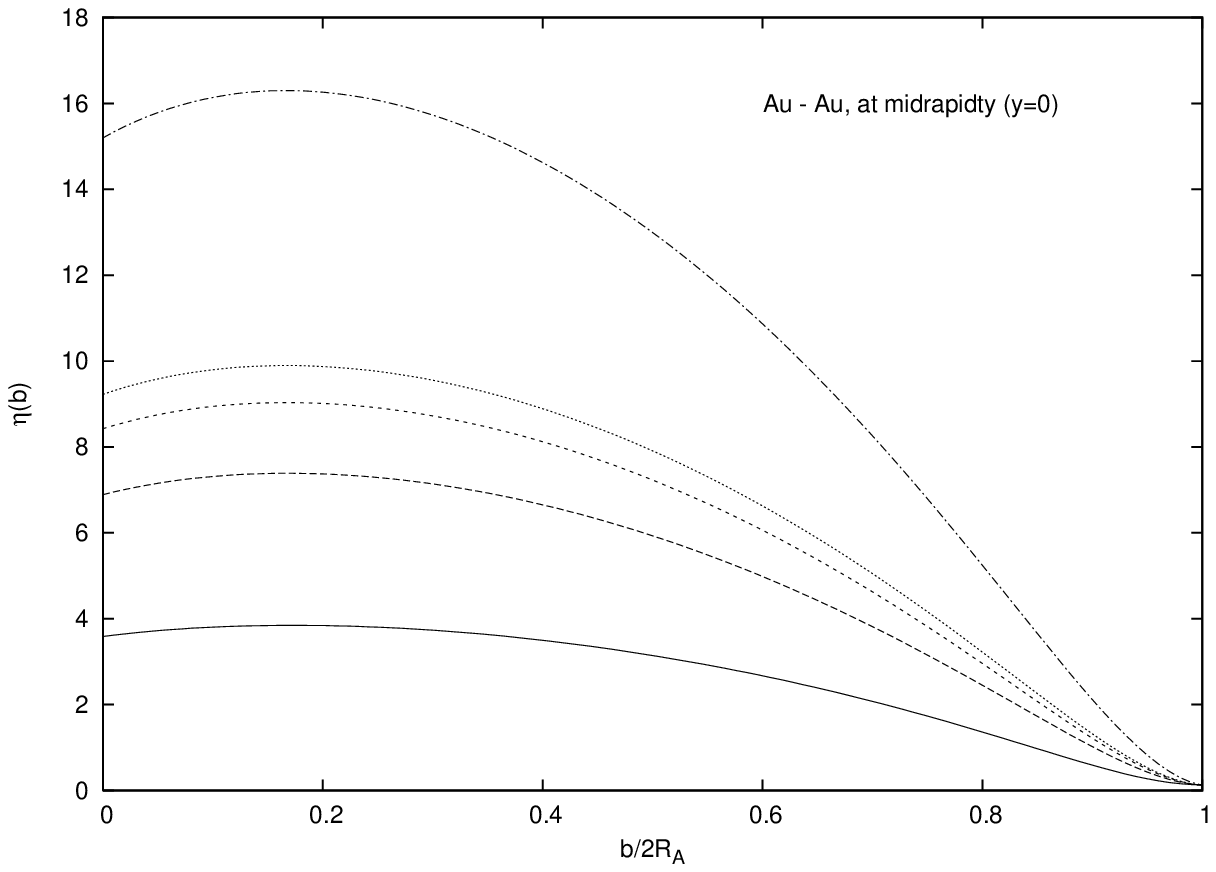}}
\caption{Percolation parameter $\eta$ in
Au-Au collisions at midrapidity as a function of impact parameter.
Curves from bottom to top correspond to energies
19.4, 62.8, 130, 200 and 6000 GeV.}
\label{Fig6}
\end{figure}
As one can observe, at RHIC and LHC
energies these values are quite large, far beyond the percolation threshold.
As a result one finds a very substantial reduction in multiplicity
calculated according to Eq. (26) with luminosities determined from
pp collisions.
In Fig. 7 and 8 we show multiplicities as a function of $y$ for
central collisions and  function of $b$ at midrapidity.
They agree rather well with the existing data both in form and absolute
values. In Fig. 8 a rather sharp change is seen in the
periphery of the nuclei, between $b/2R_A=0.8$ and unity. The RHIC data do not
exhibit such a sharp saturation. This behaviour  is a
direct consequence of using in our calculations the nuclear profile
function (23) corresponding to the  step function for the nuclear density.
A more realistic profile function would lead to a smoother transition
from the periphery to  center, although requiring a more complicated
definition of the interaction area.

Our prediction for the plateau at LHC energy (divided by
$\nu_{par}$) is around 9 for the sum of charged and neutral particles
(that is 6 for charged), which is lower than derived in other publications
~\cite{Ref13}-\cite{Ref17}. The plateau height for charged particles
in central Au-Au collisions
as a function of energy is illustrated  in Fig. 9. Of course
our results are directly related to the chosen
string radius and go upward if it is lowered. However then we loose the
agreement with the existing data.

Note that our calculations also reproduce quite well the behaviour in the
fragmentation region (limiting fragmentation). For
this the dependence of the string luminosity on energy proved to be
quite important. Without it the nice linear dependence of
multiplicities in the fragmentation region is spoiled and the line is
widened into a band.

To clearly see the effect of string fusion in Fig 10 we show the
multiplicities at $b=0$ without fusion. Their values are several times
greater than with fusion and do not agree with the experimental data at all.

Passing to the $p_T$ distributions in Fig. 11  we show
the ratios $R_{AA}(p_T)$ for central collisions at midrapidity.
The curve for  200 GeV served to determine our parameters $a,b,c$ and
$d$ in (30) and (32).
Our predictions for the LHC energy show a behaviour similar to the
RHIC energy with a still more pronounced quenching effect.
In the fragmentation region we prefer to show
the ratios $R^{par}_{AA}$ with normalization respective
to the number of participants, since in this region the
multiplicities are roughly proportional to $\nu_{par}$ due to
low string densities. Fig. 12 shows that these ratios are close
to unity and  may only fall a little below unity at LHC energies.

\section{Conclusions}
In the framework of percolation of strings we have obtained a
strong reduction of multiplicities at LHC, much larger than the rest of
models but in agreement with the extrapolation from the SPS and
RHIC experimental data. Due to  similarities between percolation of strings
and saturation of partons, it would be interesting to explore the possibility
for further reduction of multiplicities in the saturation approach.
In order to describe the energy
dependence of multiplicities in AA collisions
we need a rather large transverse size of the elementary string 0.3 fm.
This  enhances the interaction of strings and so cluster
formation, which leads to  stronger reduction of multiplicities.
In our calculations
we have used the standard optical approximation  to compute the numbers of
participants and collisions. This approximation enhances the
number of collisions for peripheral collisions in comparison with Monte-Carlo
evaluations and leads to some uncertainties also for central collisions.
For this
reason, our results must be regarded to have an uncertainty in the range of
10\%-15\% . So the string transverse size
can be lower if the number of strings is in fact lower.

Taking limiting fragmentation scaling for pp collisions as an input,
we have found  the same behaviour for AA collisions, which is
confirmed experimentally up to the RHIC energies. Our calculations
predict that limiting fragmetation scaling also remains approximately valid
at the LHC energy
(with a 5\% suppression compared with to SPS or RHIC, see Fig. 7).

We have been able to describe
reasonably well the  high transverse momentum spectrum
at different energies ranging from SPS to RHIC.
A large suppression is predicted for LHC.
In order to obtain such an agreement, in addition to the usual effects of
string clustering, such as reduction of the effective number of
independent color sources
and suppression of transverse momentum fluctuations, we need a shift of the
$p_T$ spectrum due to  energy loss. Considering string fusion
as an initial state effect (before particle production),
a final state effect is needed to account for the observed suppression,
similarly to jet quenching in the QCD picture.
\section{Acknowledgments}
This work has been partially done under contracts FPA2005-01963of Spain,
and PGIDIT03PX1 of Galicia, and also supported by the NATO grant
PST.CLG.980287 and Education Ministry of Russia grant RNP 2.1.1.1112.

%
\begin{figure}[ht]
\centerline{\epsfbox{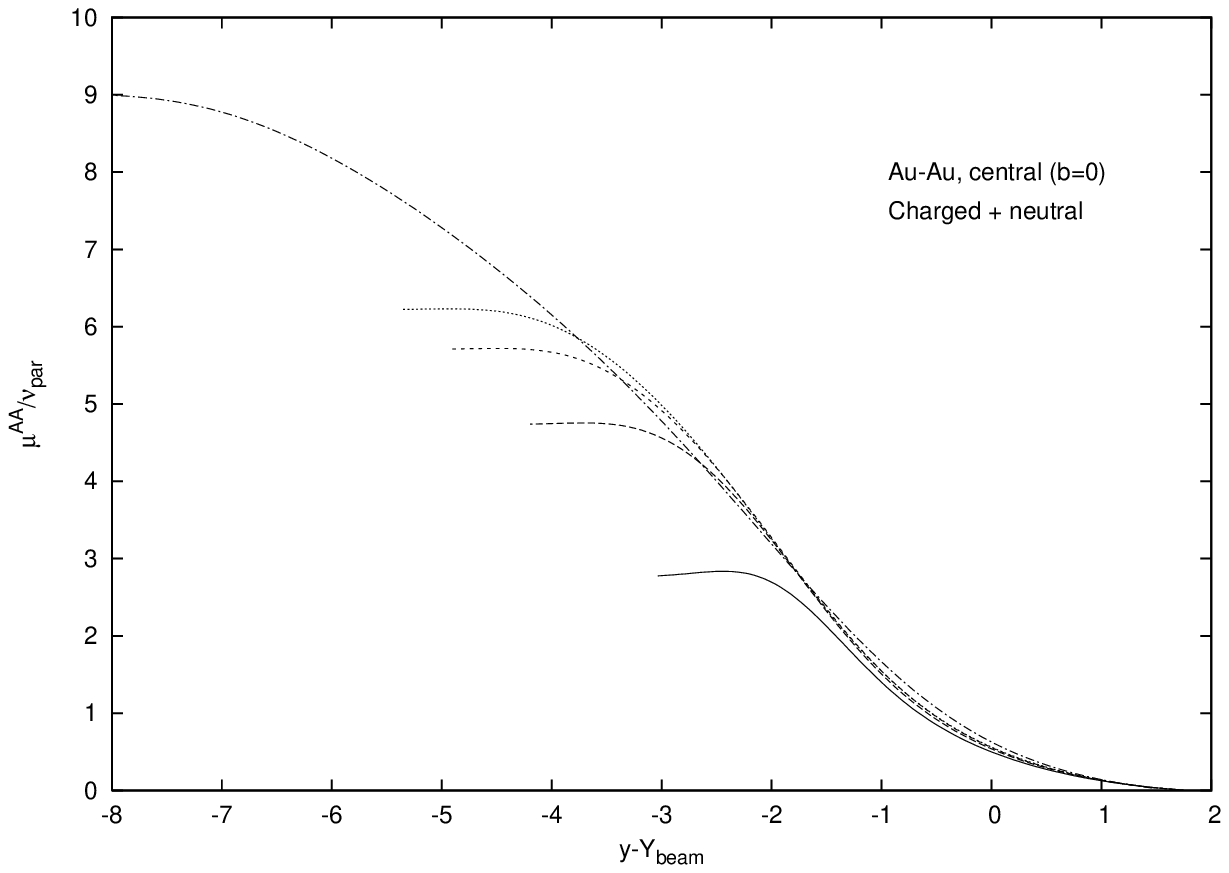}}
\caption{Multiplicities (charged plus neutrals)
per 1/2 number of participants in central
Au-Au collisions as a function of rapidity.
Curves from bottom to top correspond to energies
19.4, 62.8, 130, 200 and 6000 GeV.}
\label{Fig7}
\end{figure}
\begin{figure}[ht]
\centerline{\epsfbox{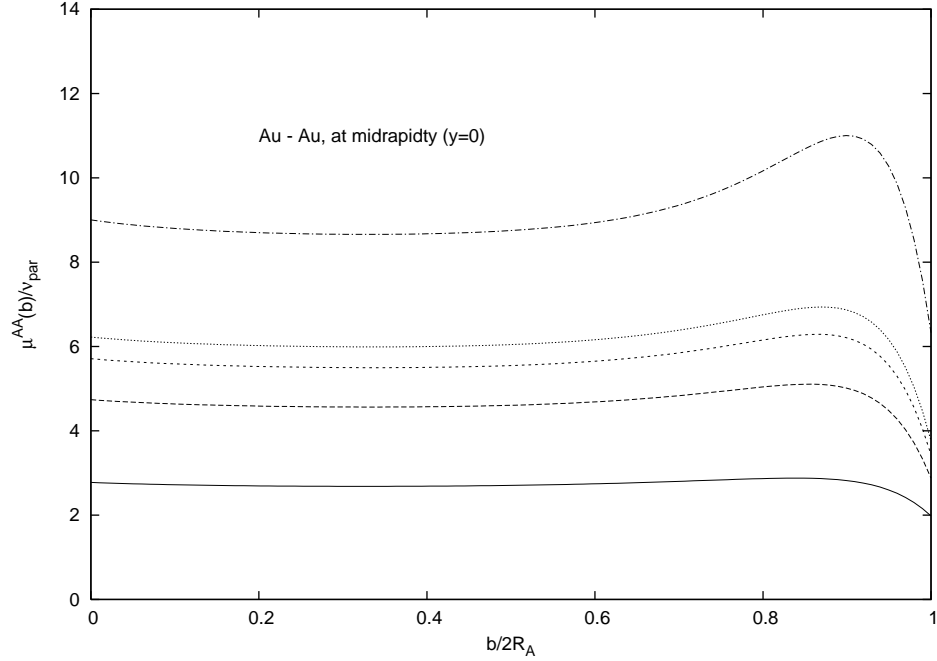}}
\caption{Multiplicities (charged plus neutrals)
per 1/2 number of participants in
Au-Au collisions at midrapidity as a function of
impact parameter.
Curves from bottom to top correspond to energies
19.4, 62.8, 130, 200 and 6000 GeV.}
\label{Fig8}
\end{figure}
\begin{figure}[ht]
\centerline{\epsfbox{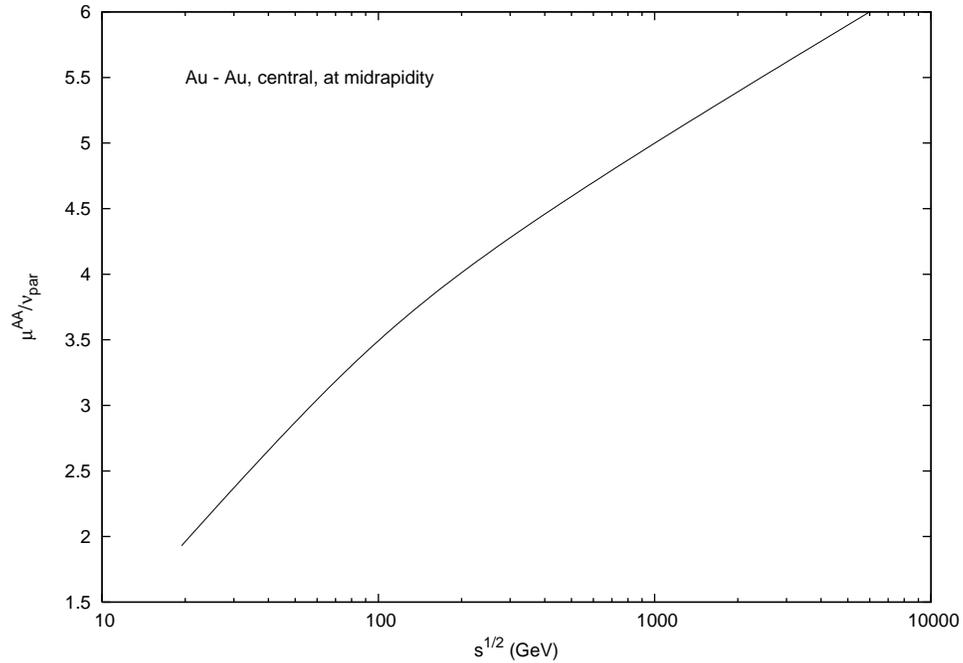}}
\caption{The plateau height for charged particles
per 1/2 number of participants in
Au-Au collisions at midrapidity as a function of
energy}
\label{Fig9}
\end{figure}
\begin{figure}[ht]
\centerline{\epsfbox{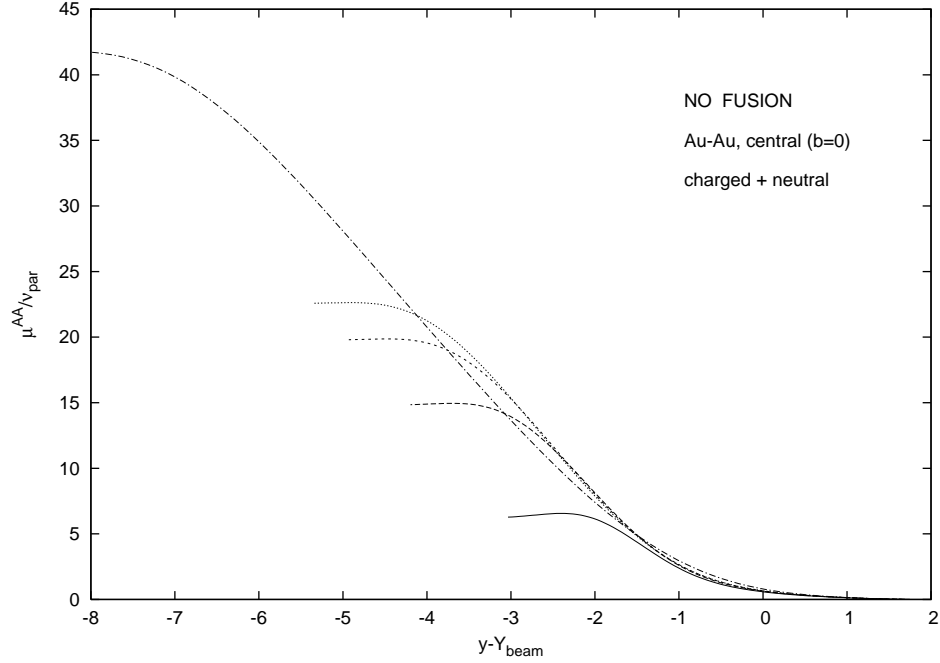}}
\caption{Multiplicities (charged plus neutrals)
per 1/2 number of participants in central
Au-Au collisions at midrapidity without fusion.
Curves from bottom to top correspond to energies
19.4, 62.8, 130, 200 and 6000 GeV.}
\label{Fig10}
\end{figure}
\begin{figure}[ht]
\centerline{\epsfbox{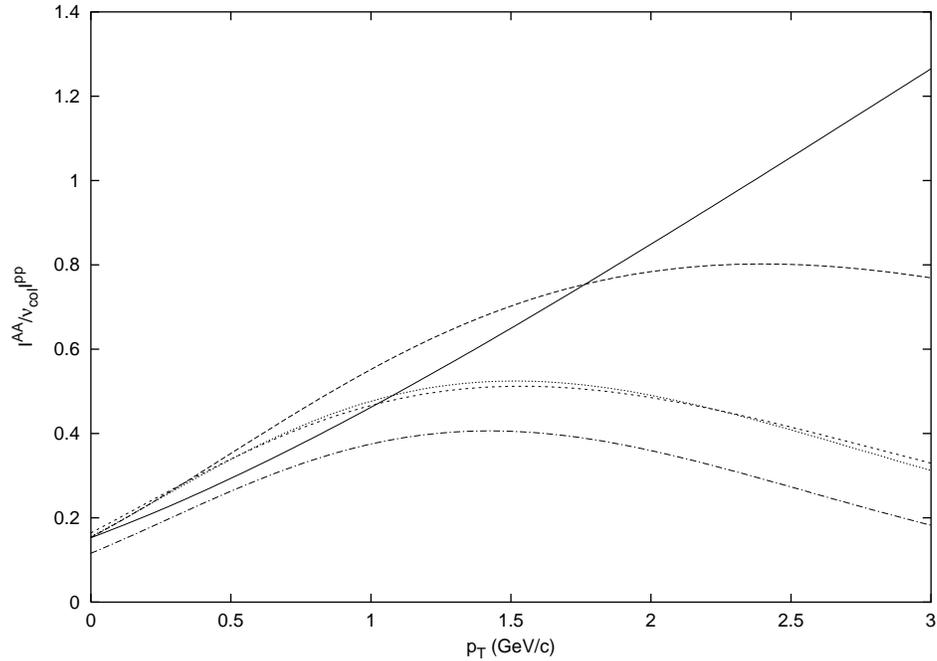}}
\caption{Nuclear factor $R_{AA}$ for central
Au-Au collisions at midrapidity.
Curves from top to bottom correspond to energies
19.4, 62.8, 130, 200 and 6000 GeV. }
\label{Fig11}
\end{figure}
\begin{figure}[ht]
\centerline{\epsfbox{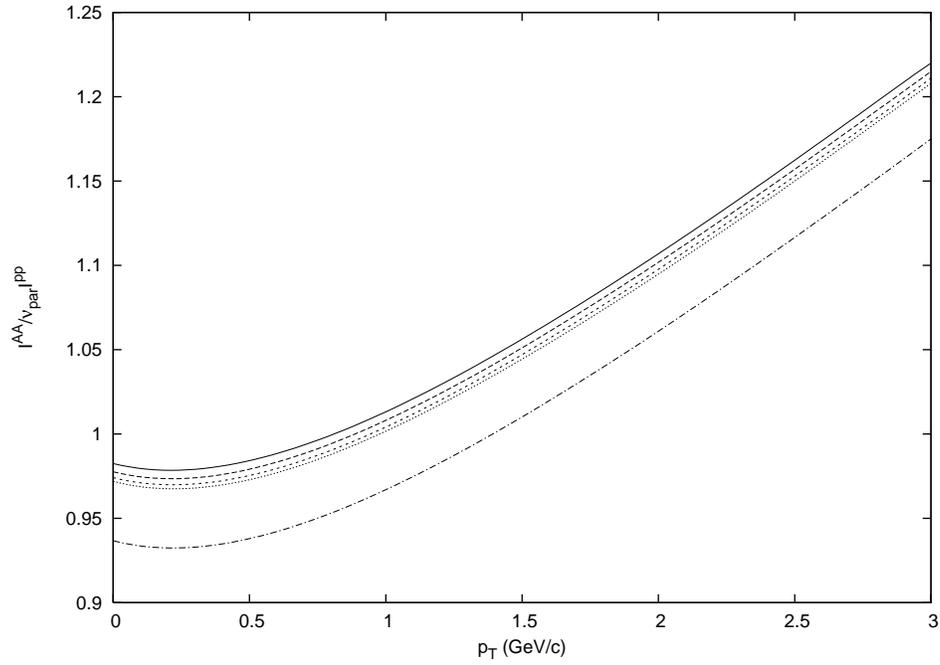}}
\caption{Nuclear factor $R^{par}_{AA}$ for central
Au-Au collisions in the fragmentation region $y=0.1Y$.
Curves from top to bottom correspond to energies
19.4, 62.8, 130, 200 and 6000 GeV. }
\label{Fig12}
\end{figure}
\end{document}